\documentclass[preprint,aps,superscriptaddress,nofootinbib]{revtex4-1}
\usepackage[T1]{fontenc}
\usepackage[latin9]{inputenc}
\usepackage{color}
\usepackage{amsmath}
\usepackage{amssymb}
\usepackage{graphicx}
\usepackage{esint}

\makeatletter
 
 \@ifundefined{textcolor}{}
 {%
   \definecolor{BLACK}{gray}{0}
   \definecolor{WHITE}{gray}{1}
   \definecolor{RED}{rgb}{1,0,0}
   \definecolor{GREEN}{rgb}{0,1,0}
   \definecolor{BLUE}{rgb}{0,0,1}
   \definecolor{CYAN}{cmyk}{1,0,0,0}
   \definecolor{MAGENTA}{cmyk}{0,1,0,0}
   \definecolor{YELLOW}{cmyk}{0,0,1,0}
 }

\usepackage{color}
\usepackage{braket}

\newcommand{\Ftwo}{\mu_{\nu_{\textrm{sa}}}}

\@ifundefined{showcaptionsetup}{}{%
 \PassOptionsToPackage{caption=false}{subfig}}
\usepackage{subfig}
\makeatother

\begin{document}

\title{Atomic ionization by sterile-to-active neutrino conversion and constraints
on dark matter sterile neutrinos with germanium detectors}

\author{Jiunn-Wei Chen}

\email{jwc@phys.ntu.edu.tw}

\affiliation{Department of Physics and Center for Theoretical Sciences, National
Taiwan University, Taipei 10617, Taiwan}

\affiliation{Leung Center for Cosmology and Particle Astrophysics, National Taiwan
University, Taipei 10617, Taiwan}

\affiliation{Center for Theoretical Physics, Massachusetts Institute of Technology,
Cambridge, MA 02139, USA}

\author{Hsin-Chang Chi}

\affiliation{Department of Physics, National Dong Hwa University, Shoufeng, Hualien
97401, Taiwan}

\author{Shin-Ted Lin}

\affiliation{Institute of Physics, Academia Sinica, Taipei 11529, Taiwan}

\affiliation{College of Physical Science and Technology, Sichuan University, Chengdu
610064, China}

\author{C.-P. Liu}

\email{cpliu@mail.ndhu.edu.tw}

\affiliation{Department of Physics, National Dong Hwa University, Shoufeng, Hualien
97401, Taiwan}

\author{Lakhwinder Singh}

\affiliation{Institute of Physics, Academia Sinica, Taipei 11529, Taiwan}

\affiliation{Department of Physics, Banaras Hindu University,Varanasi 221005,
India}

\author{Henry T. Wong}

\affiliation{Institute of Physics, Academia Sinica, Taipei 11529, Taiwan}

\author{Chih-Liang Wu}

\affiliation{Department of Physics and Center for Theoretical Sciences, National
Taiwan University, Taipei 10617, Taiwan}

\affiliation{Institute of Physics, Academia Sinica, Taipei 11529, Taiwan}

\affiliation{Center for Theoretical Physics, Massachusetts Institute of Technology,
Cambridge, MA 02139, USA}

\author{Chih-Pan Wu}

\email{jpw750811@gmail.com}

\affiliation{Department of Physics and Center for Theoretical Sciences, National
Taiwan University, Taipei 10617, Taiwan}

\affiliation{Institute of Physics, Academia Sinica, Taipei 11529, Taiwan}

\preprint{NCTS-ECP/1502, MIT-CTP/4764\textcolor{red}{{} }}
\begin{abstract}
The transition magnetic moment of a sterile-to-active neutrino conversion
gives rise to not only radiative decay of a sterile neutrino, but
also its non-standard interaction (NSI) with matter. For sterile neutrinos
of keV-mass as dark matter candidates, their decay signals are actively
searched for in cosmic X-ray spectra. In this work, we consider the
NSI that leads to atomic ionization, which can be detected by direct
dark matter experiments. It is found that this inelastic scattering
process for a nonrelativistic sterile neutrino has a pronounced enhancement
in the differential cross section at energy transfer about half of
its mass, manifesting experimentally as peaks in the measurable energy
spectra. The enhancement effects gradually smear out as the sterile
neutrino becomes relativistic. Using data taken with germanium detectors
that have fine energy resolution in keV and sub-keV regimes, constraints
on sterile neutrino mass and its transition magnetic moment are derived
and compared with those from astrophysical observations. 
\end{abstract}
\maketitle

\section{Introduction}

Sterile neutrinos are an interesting topic in particle physics, astrophysics,
and cosmology. They are invoked in many beyond-the-Standard-Model
theories, in addition to the three known ``active'' neutrinos, to
address fundamental questions including explaining the origin of neutrino
masses (a la various types of seesaw mechanisms), providing suitable
dark matter candidates (for their being massive and non-interacting
with the known forces), and setting the stage of leptogenesis that
subsequently leads to the baryon asymmetry of the universe.

As sterile neutrinos are singlets in the Standard Model gauge groups,
their masses, mixing angles, and couplings are unknown a priori --
in spite of different models having their own preferred parameter
spaces by construction, and can only be constrained by experiments.
The type I seesaw mechanism~\cite{Minkowski:1977sc,Ramond:1979py,Mohapatra:1979ia,Yanagida:1980xy}
provides a nice explanation for the smallness of active neutrino masses,
however, the predicted sterile neutrino masses at the GUT scale make
their experimental confirmation extremely difficult. On the other
hand, there are other low-energy seesaw mechanisms that predict the
existence of light sterile neutrinos (for a recent, comprehensive
overview, see the community white paper~\cite{Abazajian:2012ys}).
Two particularly interesting cases have their lightest sterile neutrino
masses~\footnote{We adopt the natural units $\hbar=c=1$ in this paper, so mass and
energy have the same dimension.} to be (i) eV scale (see, e.g., Refs.~\cite{deGouvea:2005er,Kopp:2011qd}),
which is motivated by the LSND~\cite{Aguilar:2001ty} and reactor~\cite{Mention:2011rk}
anomalies etc., and (ii) keV scale (see, e.g., Refs.~\cite{Asaka:2005an,Asaka:2005pn}),
which can be a good dark matter candidate~\cite{Dodelson:1993je,Shi:1998km,Dolgov:2000ew,Abazajian:2001nj}.
In this work, we shall focus on the latter case.

Among various venues to look for keV-scale sterile neutrinos, their
radiative decay -- if exists via the mixing of sterile (massive) and
active (light) neutrinos, as first studied in Ref.~\cite{Pal:1981rm},
or other mechanisms -- is considered one of the golden modes, in particular
by tracing the decay photons in X-ray spectra of astrophysical objects
such as galaxies or galaxies clusters~\cite{Abazajian:2001vt,Boyarsky:2006fg}.
Recently there are two groups reporting abnormal X-ray emission lines,
one at $E=(3.55-3.57)\pm0.03\,\textrm{keV}$ in a stacked XMM-Newton
spectrum of 73 galaxy clusters~\cite{Bulbul:2014sua}, and the other
at $E=3.52\pm0.02\,\textrm{keV}$ in spectra of the Andromeda galaxy
and the Perseus galaxy cluster~\cite{Boyarsky:2014jta}. These two
papers triggered a huge amount of subsequent theoretical interpretations,
and the possibility of a decaying sterile neutrino with mass $m_{s}=7.1\,\textrm{keV}$
and its mixing to active neutrino mixing with an angle $\sin^{2}(2\theta)\sim(5-7)\times10^{-11}$
was suggested~\cite{Bulbul:2014sua,Boyarsky:2014jta}. 

Motivated by this anomalous X-ray emission line and its possible interpretation
of radiative sterile neutrino decay, we consider in this paper an
effective Lagrangian that gives rise to the coupling of an incoming
sterile neutrino, an outgoing active neutrino, and a virtual photon.
When the virtual photon couples to electromagnetic currents of normal
matter, this will generate some nonstandard interaction between the
sterile neutrino and normal matter and the signals, in forms of normal
matter recoils, can be searched for with typical direct dark matter
detectors. In this way, direct detectors can provide complementary
constraints on sterile neutrino properties to the above indirect astrophysical
searches in X-ray spectra, which originate from the same effective
Lagrangian with the photon becomes real, outgoing, thus observable.

The specific process we study is atomic ionization caused by the electromagnetic
field generated during the sterile-to-active neutrino conversion.
Effectively, this field is due to a transition magnetic moment that
a sterile neutrino possesses, either by oscillation to active species
or other new-physics mechanisms. One interesting feature of this process
is the exchanged photon can go across the space-like region (typical
for $t$-channel processes) to the time-like region (e.g., the final
active, massless neutrino has almost zero four momentum so the square
of 4-momentum transfer is simply $\sim m_{s}^{2}$), because the two-body
atomic final state (ion plus ionized electron) can make this kinematically
possible. As a result, there is a big enhancement from the photon
pole in the cross section.

For keV-mass, nonrelativistic, sterile neutrinos, low-energy detectors
with capability of sub-GeV thresholds are required, so we focus on
germanium detectors and use their data to set constraints on sterile
neutrino properties. We also note that there are related studies,
for example, the active-to-sterile neutrino conversion in magnetic
environments of the early universe~\cite{Semikoz:1994uy} and supernovae~\cite{Sahu:1998jh};
sterile neutrino production the by Primakoff effect in neutrino beams~\cite{Gninenko:1998nn};
and implications of sterile neutrinos in searches of magnetic moments
of active neutrinos~\cite{Balantekin:2013sda}.

The paper is organized as follows. In Sec.~\ref{sec:Formalism},
we lay down the basic formalism that describes the atomic ionization
caused by the transition magnetic moment of a sterile neutrino, and
emphasize the differences from the more familiar cases where incoming
and outgoing neutrinos are nearly mass-degenerate. In Secs.~\ref{sec:Hydrogen}
and \ref{sec:Germanium}, we present and discuss our results for hydrogen
and germanium ionization, respectively. Using data taken by germanium
detectors, we derive the bounds on sterile neutrino properties in
Sec.~\ref{sec:Bounds}, and conclude in Sec.~\ref{sec:Summary}.

\section{Formalism \label{sec:Formalism}}

The radiative decay of a sterile neutrino $\nu_{s}$ into a Standard-Model
neutrino $\nu_{a}$, $\nu_{s}\rightarrow\nu_{a}+\gamma$, can be effectively
formulated by the interaction Lagrangian 
\begin{equation}
\mathcal{L}_{\nu_{s}\nu_{a}\gamma}=\frac{\Ftwo e}{2m_{e}}\frac{1}{2}\bar{\nu}_{a}\sigma_{\mu\nu}\nu_{s}F^{\mu\nu}\,,\label{eq:L_nu_s-n_a-g}
\end{equation}
where $\sigma_{\mu\nu}=\frac{i}{2}[\gamma_{\mu}\,,\,\gamma_{\nu}]$
is the tensor Dirac matrix and $F^{\mu\nu}=\partial^{\mu}A^{\nu}-\partial^{\nu}A^{\mu}$
is the field strength of the photon field $A^{\mu}$. The coupling
constant $\Ftwo$ can be understood as the transition magnetic moment
measured in units of Bohr's magneton, $e/(2m_{e})$, that induces
a $\nu_{s}$--$\nu_{a}$ transition. In the limit that $\nu_{a}$
is massless, the total decay width can thus be expressed as 
\begin{equation}
\Gamma_{\nu_{s}\rightarrow\nu_{a}\gamma}=\left(\frac{\Ftwo e}{2m_{e}}\right)^{2}\frac{m_{s}^{3}}{8\pi}\,.\label{eq:nu_s_decay}
\end{equation}

In case $\nu_{s}$ oscillates into $\nu_{a}$ by a mixing angle $\theta$,
the transition magnetic moment can arise from the one-loop radiative
corrections. As calculated in Ref~\cite{Pal:1981rm}, 
\begin{equation}
\Ftwo=\frac{3G_{F}}{4\sqrt{2}\pi^{2}}m_{e}m_{s}\sin\theta\,,
\end{equation}
where $G_{F}$ is the Fermi constant, this leads to the familiar expression
that relates $\Gamma_{\nu_{s}\rightarrow\nu_{a}\gamma}$ to the oscillation
angle of sterile neutrinos 
\begin{equation}
\Gamma_{\nu_{s}\rightarrow\nu_{a}\gamma}^{(\textrm{osc.})}=\frac{9}{1024\pi^{4}}G_{F}^{2}\,\alpha\,\sin^{2}(2\theta)\,m_{s}^{5}\,,
\end{equation}
where $\alpha=e^{2}/(4\pi)$ is the fine structure constant. 

With $\mathcal{L}_{\nu_{s}\nu_{a}\gamma}$, it is also possible to
consider processes where a $\nu_{s}$ is converted to a $\nu_{a}$
by scattering off an electromagnetic source, where the exchanged photon
becomes virtual. In this article, we are interested in atomic ionization,
i.e., 
\[
\nu_{s}+\textrm{A}\rightarrow\nu_{a}+\textrm{A}^{+}+e^{-}\,,
\]
because the recoiled electron can be detected as a signal of such
$\nu_{s}$--$\nu_{e}$ conversion. Since this process resemble the
one 
\[
\nu_{a}+\textrm{A}\rightarrow\nu_{a}+\textrm{A}^{+}+e^{-}\,,
\]
which is used to constrain the magnetic moments of the Standard-Model
neutrinos~\cite{Chen:2013iud}, except for different mass and kinematics
of the incident neutrinos, we shall not repeat a full derivation of
the scattering formalism but only highlight the main result with key
differences. 

The single differential cross section with respect to the energy transfer
$T$ by neutrinos is expressed in the form 
\begin{equation}
\dfrac{d\sigma}{dT}=\int d\cos\theta\,\dfrac{2\pi\alpha^{2}}{(q^{2})^{2}}\,\dfrac{|\vec{k}_{a}|}{|\vec{k}_{s}|}\left(\frac{\Ftwo}{2m_{e}}\right)^{2}(V_{L}R_{L}+V_{T}R_{T})\,,\label{eq:ds/dT_bare}
\end{equation}
where $\vec{k}_{s}$ and $\vec{k}_{a}$ are the 3-momenta of the incoming
and outgoing neutrinos, respectively; $q^{2}=q_{\mu}q^{\mu}$ is the
square of the four momentum transfer $q_{\mu}=(T,\vec{q})$; and the
integration over the scattering angle of neutrino, $\theta$, is confined
in the range: 
\begin{equation}
\min\left\{ 1,\max\left[-1,\dfrac{\vec{k}_{s}^{2}+\vec{k}_{a}^{2}-2M(T-B)}{2|\vec{k}_{s}||\vec{k}_{a}|}\right]\right\} \leq\cos\theta\leq1\,,\label{eq:cos_th}
\end{equation}
with $M=m_{\textrm{A}^{+}}+m_{e}$ being the total mass of the final
$\textrm{A}^{+}+e^{-}$ system and $B$ the binding energy of the
ejected electron.

The unpolarized longitudinal and transverse response functions, $R_{L}$
and $R_{T}$, which are functions of $T$ and $|\vec{q}|$, are defined
by 
\begin{eqnarray}
R_{L} & = & \sum_{f}\overline{\sum_{i}}|\left<f|\rho^{(A)}(\vec{q})|i\right>|^{2}\,\delta\left(T-B-\dfrac{\vec{q}^{2}}{2M}-\dfrac{\vec{p}_{r}^{2}}{2\mu}\right)\,,\\
R_{T} & = & \sum_{f}\overline{\sum_{i}}|\left<f|\vec{j}_{\perp}^{(A)}(\vec{q})|i\right>|^{2}\,\delta\left(T-B-\dfrac{\vec{q}^{2}}{2M}-\dfrac{\vec{p}_{r}^{2}}{2\mu}\right)\,.
\end{eqnarray}
The former depends on the atomic charge density $\rho^{(A)}(\vec{q})$
and the latter on the atomic transverse (perpendicular to $\vec{q}$)
current density $\vec{j}_{\perp}^{(A)}(\vec{q})$. For all processes
we are going to discuss in this paper, the nuclear charge and current
are negligible. Note that the spin states of the initial atom $\ket{i}$
are averaged (hence the symbol $\overline{\sum}_{i}$) and the final
states $\ket{f}$ are summed (integrated if quantum numbers are continuous).
The delta function imposes the energy conservation, and the total
energy of the final $\textrm{A}^{+}+e^{-}$ system is separated into
the center-of-mass part $\vec{q}^{2}/(2M)$ and the internal part
$\vec{p}_{r}^{2}/(2\mu)$, where $\vec{p}_{r}$ is the relative momentum
and the reduced mass $\mu=m_{e}m_{\textrm{A}^{+}}/M\approx m_{e}$.

Because the incoming and outgoing neutrinos have different masses,
the kinematic factors $V_{L}$ and $V_{T}$ now become 
\begin{align}
V_{L} & =\dfrac{-q^{4}}{|\vec{q}|^{4}}\bigg((m_{s}+m_{a})^{2}|\vec{q}|^{2}+(E_{s}+E_{a})^{2}q^{2}+(m_{s}^{2}-m_{a}^{2})^{2}-2(E_{s}^{2}-E_{a}^{2})(m_{s}^{2}-m_{a}^{2})\bigg)\,,\label{eq:V_L}\\
V_{T} & =-\Big((m_{s}+m_{a})^{2}-\dfrac{(m_{s}^{2}-m_{a}^{2})^{2}}{q^{2}}\Big)q^{2}-\dfrac{q^{2}\Big(4|k_{s}|^{2}|k_{a}|^{2}-(|k_{s}|^{2}+|k_{a}|^{2}-|\vec{q}|^{2})^{2}\Big)}{2|\vec{q}|^{2}}\,.\label{eq:V_T}
\end{align}
If one set $m_{s}=m_{a}=m_{l}$, then the above results converge to
Eqs.~(13,14) of Ref.~\cite{Chen:2013iud}. When weighted by the
factor $1/(q^{2})^{2}$ -- which arises from the square of photon
propagator -- in calculating the differential cross section as Eq.~(\ref{eq:ds/dT_bare}),
the $V_{T}/q^{4}$ term yields a double pole in $q^{2}$ with the
coefficient $(m_{s}^{2}-m_{a}^{2})^{2}$ and the rest being a single
pole in $q^{2}$, while $V_{L}/q^{4}$ term does not contain any pole
in $q^{2}$. In cases where $q^{2}=0$ is never reached, for example,
elastic scattering $\nu_{s}+\textrm{A}\rightarrow\nu_{a}+\textrm{A}$
in which $q^{2}=-2M_{\textrm{A}}T$, there will not be any singularity
in $d\sigma/dT$. However, the situation is quite different when the
atom is ionized into a two-body system $\textrm{A}^{+}+e^{-}$. For
simplicity, take the $m_{a}\rightarrow0$ limit so 
\begin{equation}
q^{2}=m_{s}^{2}-2E_{s}(E_{s}-T)+2(E_{s}-T)\sqrt{E_{s}^{2}-m_{s}^{2}}\cos\theta\,.\label{eq:q^2}
\end{equation}
As the scattering angle can vary in the range given by Eq.~(\ref{eq:cos_th})
(unlike the above elastic scattering case where $\cos\theta$ is fixed),
it is possible to find values of $T$ for a given $m_{s}$ such that
the $q^{2}=0$ pole is kinematically accessible. 

In order to obtain a physical, finite differential cross section,
apparently the singularity due to the real photon pole needs some
regularization. Notice that when the exchanged virtual photon approaches
the on-shell limit $q^{2}=0$, the scattering process is no longer
distinguishable from a two-step process in which the sterile neutrino
first undergoes a radiative decay then the emitted real photon subsequently
causes the atomic ionization. This two-step process causes an attenuation
of the real photon intensity in dense detector media, and can be easily
implemented by adding a small imaginary part to the wave number, i.e.,
\begin{equation}
|\vec{q}|\rightarrow|\vec{q}|+\frac{i}{2}n_{\textrm{A}}\sigma_{\gamma}(\textrm{A},|\vec{q}|)\,,
\end{equation}
where $n_{\textrm{A}}$ is the number density of scatterers $A$ in
detector and $\sigma_{\gamma}(\textrm{A},|\vec{q}|)$ is the photoabsorption
cross section of one single scatterer $A$ with photon energy $T=|\vec{q}|$.
As a result, 
\begin{equation}
|\vec{q}|^{2}\rightarrow|\vec{q}|^{2}-n_{\textrm{A}}^{2}\sigma_{\gamma}^{2}+i\,|\vec{q}|n_{\textrm{A}}\sigma_{\gamma}\approx|\vec{q}|^{2}+i\,|\vec{q}|n_{\textrm{A}}\sigma_{\gamma}(\textrm{A},|\vec{q}|)\,,
\end{equation}
and the photon propagator should be regularized as 
\[
\frac{1}{q^{2}}\rightarrow\frac{1}{q^{2}-i\,|\vec{q}|n_{\textrm{A}}\sigma_{\gamma}(\textrm{A},|\vec{q}|)}\,.
\]
Using this ansatz, the regularized differential cross section (denoted
by a bar) 

\begin{equation}
\dfrac{d\bar{\sigma}}{dT}=\int d\cos\theta\,\dfrac{2\pi\alpha^{2}}{(q^{2})^{2}+(|\vec{q}|n_{\textrm{A}}\sigma_{\gamma}(\textrm{A},|\vec{q}|))^{2}}\,\dfrac{|\vec{k}_{a}|}{|\vec{k}_{s}|}\left(\frac{\Ftwo}{2m_{e}}\right)^{2}(V_{L}R_{L}+V_{T}R_{T})\,\label{eq:ds/dT_reg}
\end{equation}
is free of singularity, and away from the pole region $q^{2}\rightarrow0$,
the regulator $|\vec{q}|n_{\textrm{A}}\sigma_{\gamma}(\textrm{A},|\vec{q}|)$
should have negligible impact.

\subsection*{Some approximation schemes}

Full calculations of $d\bar{\sigma}/dT$ require many-body wave functions
so that the response functions $R_{L}$ and $R_{T}$ can be evaluated.
In most cases, these are highly non-trivial, so we discuss in the
following a few approximation schemes that help to simplify the many-body
problems in certain, if not all, kinematic regions. 

First, when the real photon pole $q^{2}\rightarrow0$ is accessed
(or approached) in a scattering process, it is natural to expect that
the equivalent photon approximation (EPA) should work as the pole
region dominates the differential cross section: 
\begin{eqnarray}
\frac{d\bar{\sigma}}{dT}\bigg|_{\textrm{EPA}} & = & \left(\frac{\Ftwo}{2m_{e}}\right)^{2}\frac{\alpha}{\pi}\frac{|\vec{k_{a}}|}{|\vec{k_{s}}|}T\sigma_{\gamma}(\text{\textrm{A},}T)\int d\cos\theta\,\frac{V_{T}}{q^{4}+(|\vec{q}|n_{\textrm{A}}\sigma_{\gamma}(\textrm{A},|\vec{q}|))^{2}}\,.\label{eq:ds/dT_EPA}
\end{eqnarray}
In our considered process, it can be easily worked out that for energy
transfer $T$ within the interval $[(E_{s}-|\vec{k}_{s}|)/2,(E_{s}+|\vec{k}_{s}|)/2]$,
a photon pole always occurs at some certain scattering angle.~\footnote{Note that the physical range of $T$ is between $[B,E_{s}]$.} 

Under further approximations that (i) $V_{T}\approx(m_{s}^{2}-m_{a}^{2})^{2}$,
which is the most singular term in $V_{T}$ that comes with a double
pole $1/(q^{2})^{2}$, and (ii) the regulator can be set to a constant
$Tn_{\textrm{A}}\sigma_{\gamma}(\textrm{A},T)$ since it is only important
when $|\vec{q}|=T$, the integration of the EPA formula can be simplified
and yields 
\begin{eqnarray}
\frac{d\bar{\sigma}}{dT}\bigg|_{\textrm{EPA}}^{\textrm{pole}} & \approx & \left(\frac{\Ftwo}{2m_{e}}\right)^{2}\frac{\alpha}{2\pi}\frac{1}{|\vec{k_{s}}|^{2}}\dfrac{(m_{s}^{2}-m_{a}^{2})^{2}}{n_{\textrm{A}}}\left.\tan^{-1}\left(\dfrac{q^{2}}{Tn_{\textrm{A}}\sigma_{\gamma}(\textrm{A},T)}\right)\right|_{q_{\textrm{min}}^{2}}^{q_{\textrm{max}}^{2}}\,,\\
 & \approx & \left(\frac{\Ftwo}{2m_{e}}\right)^{2}\frac{\alpha}{2n_{\textrm{A}}}\frac{(m_{s}^{2}-m_{a}^{2})^{2}}{|\vec{k_{s}}|^{2}}\,,\quad\textrm{if }|q_{\textrm{max}}^{2}|\textrm{ and }|q_{\textrm{min}}^{2}|\gg Tn_{\textrm{A}}\sigma_{\gamma}(\textrm{A},T)\,.\label{eq:ds/dT_EPA_approx}
\end{eqnarray}
The last line indicates for cases where both $|q_{\textrm{max}}^{2}|$
and $|q_{\textrm{min}}^{2}|$ are much larger than the regulator $Tn_{\textrm{A}}\sigma_{\gamma}(\textrm{A},T)$,
the approximated EPA result takes an extremely simple form that is
independent of $T$, as long as $(E_{s}-|\vec{k}_{s}|)/2\le T\le(E_{s}+|\vec{k}_{s}|)/2$.
In later sections, we will give example for such plateau-like pattern
in $d\bar{\sigma}/dT$ to illustrate this point.

Second, in contrast to EPA, one can keep only the longitudinal response
$R_{L}$ by setting $V_{T}=0$ in Eq.~(\ref{eq:ds/dT_reg}). This
corresponds to the case where the exchanged photon is purely longitudinal,
so is called the longitudinal photon approximation (LPA). Depending
on the kinematics of the processes being considered, the LPA can work
well, in particular for the cases where the atomic 3-current density
is relatively suppressed than the charge density in nonrelativistic
expansion, or the exchanged photon is not close to real since there
is no $q^{2}=0$ pole in $V_{L}/q^{4}$. 

Last by not least, one can neglect the binding effect on atomic wave
functions and treat atomic electrons as free particles, as long as
the deposited energy $T$ is big enough to yield ionization. This
free electron approximation (FEA) is done by multiplying the scattering
cross section of free electrons, $d\bar{\sigma}^{(\nu e)}/dT$, by
the number of bound electrons that can be ionized with a given $T$: 

\begin{eqnarray}
\dfrac{d\bar{\sigma}}{dT}\bigg|_{\textrm{FEA}} & = & \sum_{i}\,\theta(T-B_{i})\,\dfrac{d\bar{\sigma}}{dT}^{(\nu_{s}e\rightarrow\nu_{a}e)}\,,
\end{eqnarray}
where

\begin{eqnarray}
\dfrac{d\bar{\sigma}}{dT}^{(\nu_{s}e\rightarrow\nu_{a}e)} & = & \left(\dfrac{\Ftwo}{2m_{e}}\right)^{2}\dfrac{\pi\alpha^{2}}{m_{e}|\vec{k_{s}}|^{2}}\dfrac{1}{q^{4}+(|\vec{q}|n_{\textrm{A}}\sigma_{\gamma}(\textrm{A},|\vec{q}|))^{2}}\nonumber \\
 &  & \times\bigg\{\left(q^{2}(m_{s}+m_{a})^{2}-(m_{s}^{2}-m_{a}^{2})^{2}\right)(2m_{e}^{2}+q^{2})-q^{4}(m_{s}^{2}+m_{a}^{2})\nonumber \\
 &  & -2q^{2}m_{e}(2E_{s}-T)(m_{1}^{2}-m_{2}^{2})-8q^{2}m_{e}^{2}E_{s}(E_{s}-T)\bigg\}\bigg|_{q^{2}=-2m_{e}T}\,.
\end{eqnarray}
Note that $q^{2}=-2m_{e}T<0$ in the FEA, so the real photon pole
can not be reached, and typically one has $2m_{e}T\gg|\vec{q}|n_{\textrm{A}}\sigma_{\gamma}(\textrm{A},|\vec{q}|)$,
where $\vec{q}^{2}=2m_{e}T+T^{2}$, so the regulator here is not important.

\section{Hydrogen Case \label{sec:Hydrogen}}

In this section, we consider the hydrogen atom as the target with
different combinations of sterile neutrino masses $m_{s}$ and velocities
$v_{s}$. The number density of hydrogen atoms is taken to be the
one in water, i.e., $n_{\textrm{H}}=6.6\times10^{22}/\textrm{cm}^{3}$,
and this gives the regulator $|\vec{q}|n_{\textrm{\textrm{H}}}\sigma_{\gamma}(\textrm{H},|\vec{q}|)\lesssim120\,\textrm{eV}^{2}$
in the allowed range of $|\vec{q}|$ (the regulator decreases with
increasing $|\vec{q}|$). Note that most calculations in this case
can be done analytically by standard techniques, and details be found
in Ref.~\cite{Chen:2013iud}.

\subsection{$m_{s}=7.1\,\textrm{keV}$, $v_{s}=10^{-3}$}

\begin{figure}
\subfloat[$m_{s}=7.1\,\textrm{keV}$]{\includegraphics[width=0.33\textwidth]{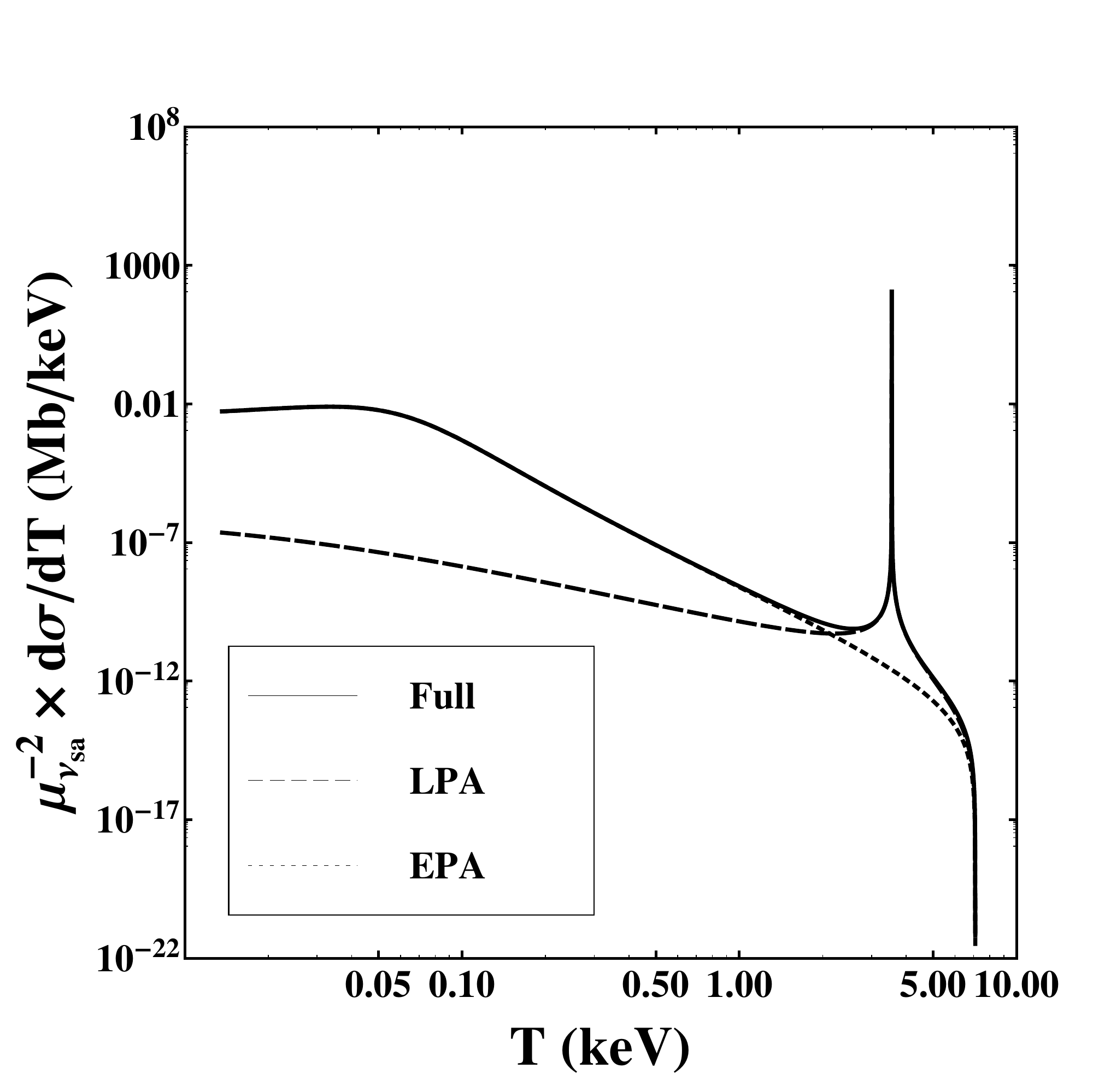}

\label{fig:nu_s-H_NR-1}

}\subfloat[$m_{s}=100\,\textrm{keV}$]{\includegraphics[width=0.33\textwidth]{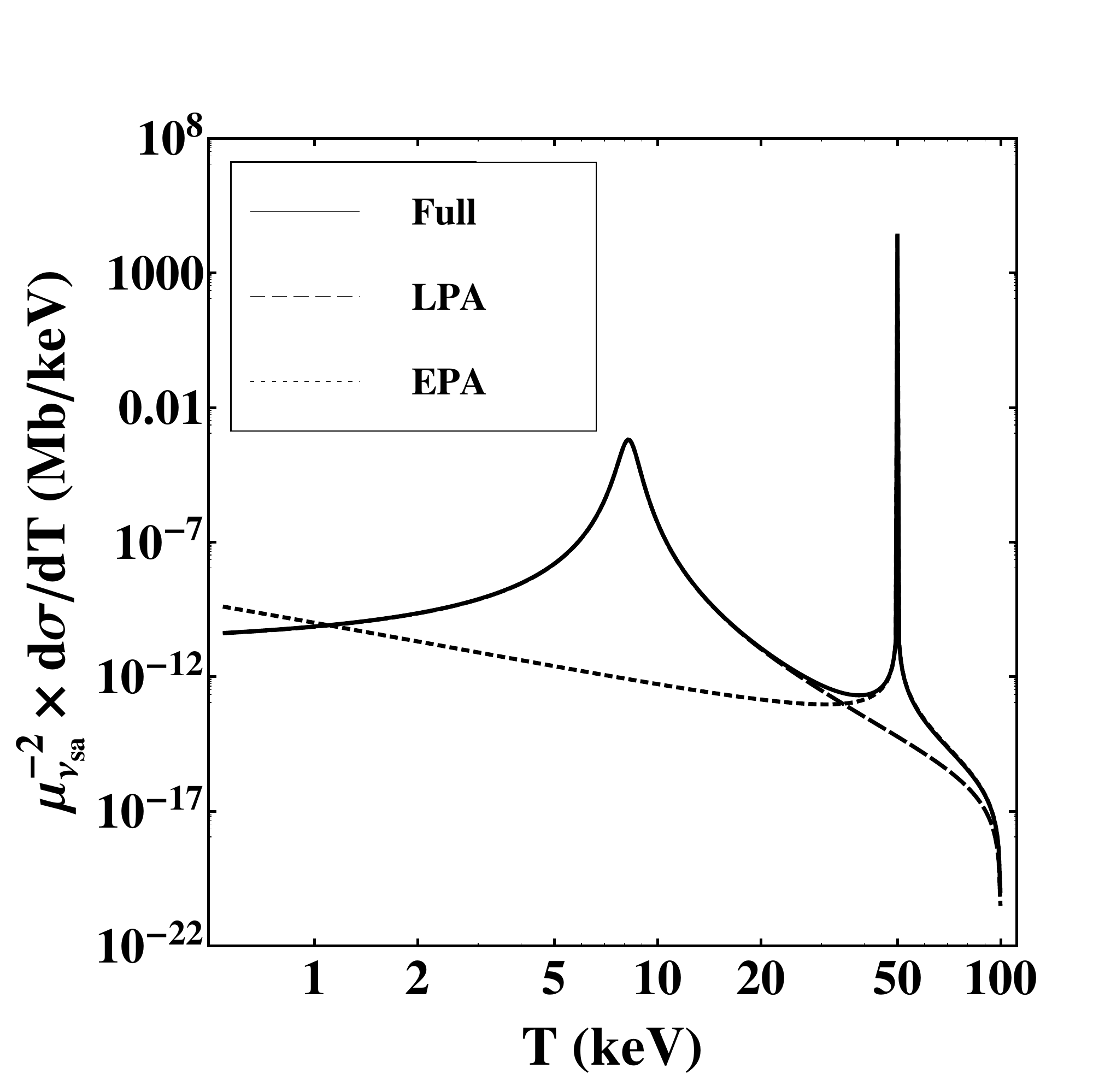}

\label{fig:nu_s-H_NR-2}

}\subfloat[$m_{s}=1\,\textrm{MeV}$]{\includegraphics[width=0.33\textwidth]{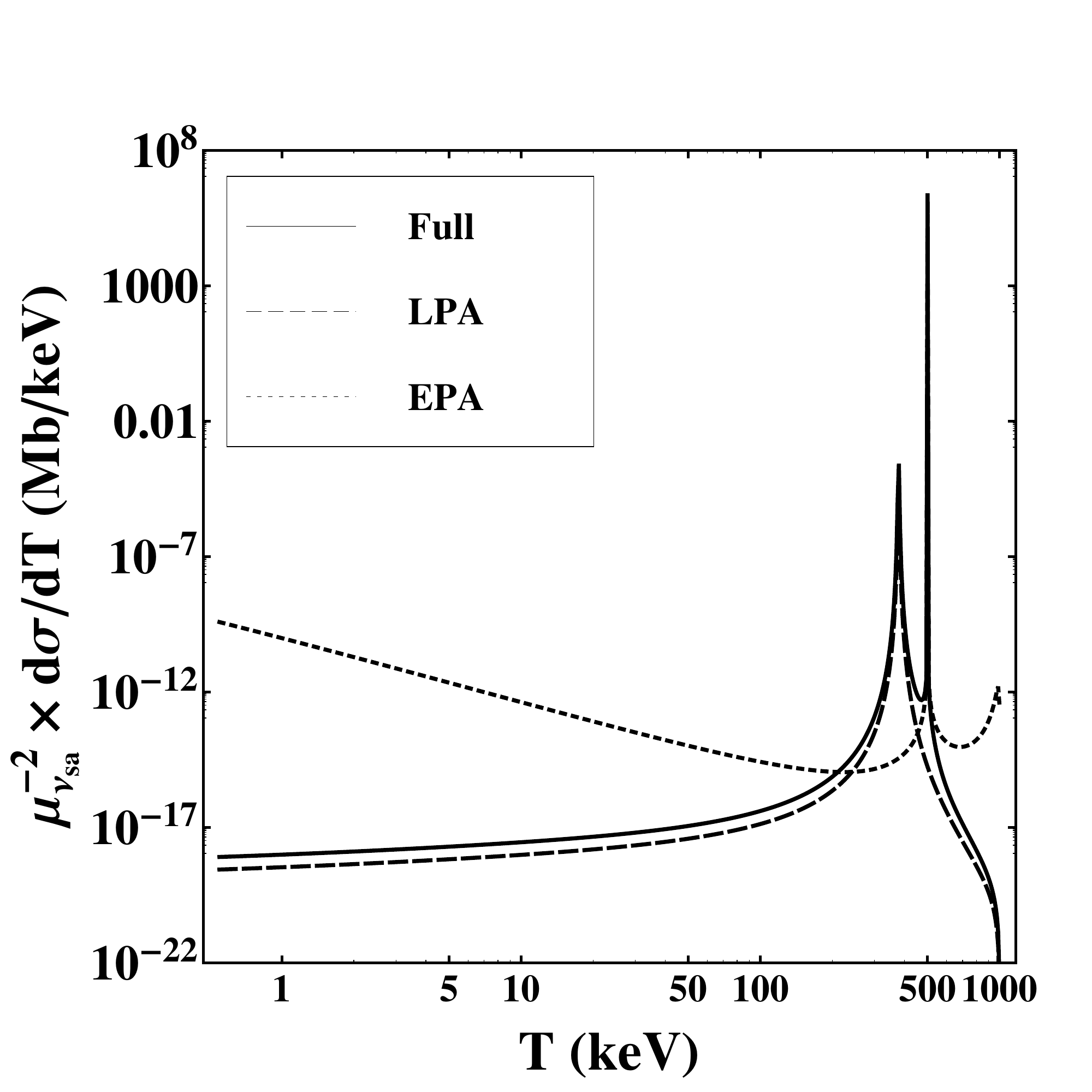}

\label{fig:nu_s-H_NR-3}

}

\caption{The differential scattering cross sections of nonrelativistic sterile
neutrinos ($v_{s}=10^{-3}$) and hydrogen atoms through the transition
magnetic moment $\Ftwo$ with selected $m_{s}$. \label{fig:nu_s-H_NR}}
\end{figure}

Suppose the X-ray anomaly hints the existence of 7.1-keV-mass sterile
neutrinos as a form of cold DM in our galaxy with NR velocity on the
order of $10^{-3}$. The differential scattering cross section of
such sterile neutrinos and hydrogen atoms through the transition magnetic
moment $\Ftwo$ is shown in Fig.~\ref{fig:nu_s-H_NR-1}. As can be
seen from the plot, the EPA works well around the near-pole region,
i.e., $T\sim m_{s}/2$, while the rest part is better approximated
by the LPA because the longitudinal response dominates herein. The
sharp peak around the $T\sim m_{s}/2$ pole region therefore gives
rise to an much enhanced sensitivity to $F_{2}^{2}$ -- if detectors
have good energy resolution.

\subsection{$m_{s}=100\,\textrm{keV};\,1\,\textrm{MeV}$ , $v_{s}=10^{-3}$}

Consider the mass of the NR sterile neutrino is increased to 100 keV
and 1 MeV, the differential cross sections are shown in Figs.~\ref{fig:nu_s-H_NR-2}
and \ref{fig:nu_s-H_NR-3}, respectively. Unlike the previous case
with $m_{s}=7.1\,\textrm{keV}$, $d\sigma/dT$ now exhibits a twin-peak
pattern. The peak at $T=m_{s}/2=50\,\textrm{keV}$ or $500\,\textrm{keV}$
is the one that is due to the double pole in the photon propagator,
so can be well approximated by the EPA. Compared with the peak in
the $m_{s}=7.1\,\textrm{keV}$ case, not only its absolute value is
bigger because of a larger $m_{s}$ {[}see Eq.~(\ref{eq:ds/dT_EPA_approx}){]}
but also it stands out more significantly from the rest non-peak region,
where the atomic longitudinal response is more suppressed for a bigger
momentum transfer leads to a more oscillating integration result. 

The other peak at $T\sim8.18\,\textrm{keV}$ or $333\,\textrm{keV}$
can be understood in the following way: Since it is well described
by the LPA, this implies the longitudinal response, $R_{L}$, dominates
in this region. The maximum of $R_{L}$ is reached under the condition:
$|\vec{q}|\sim|\vec{p_{r}}|\approx|\vec{p_{e}}|$; in other words,
the momentum (and energy, too) transfer is purely taken by the electron,
while the proton is just a spectator. In such case, the scattering
appears to be a two-body process so that $q^{2}\approx-2m_{e}T$ and
$\vec{q}^{2}\approx(m_{s}-T)^{2}$. As a result, the energy transfer
that gives rise to this peak due to two-body kinematics is 

\begin{equation}
T^{(\nu e)}=\dfrac{m_{s}^{2}}{2(m_{s}+m_{e})}\,.\label{eq:T_nu-e}
\end{equation}
For $m_{s}=7.1\,\textrm{keV}$, one would predict a similar peak happening
at $T^{(\nu e)}=48.6\,\textrm{keV}$; in fact this can be readily
seen in Fig.~\ref{fig:nu_s-H_NR-1} but without a sharp contrast,
for kinematic reason just discussed.

\subsection{$m_{s}=7.1\,\textrm{keV}$, $v_{s}\rightarrow1$}

As there might be possible mechanisms to boost cold DM candidates,
it is also interesting to consider relativistic 7.1-keV-mass sterile
neutrinos. For this case, we need to discuss first the broadening
effect in decay and scattering of boosted sterile neutrinos. 

Suppose the sterile neutrino moves in some relativistic velocity $v_{s}\rightarrow1$.
In the rest frame of the sterile neutrino, when it breaks up into
a photon and a light neutrino (taken to be massless in our consideration),
the energy of the photon has a single value $m_{s}/2$. After transforming
back to the laboratory frame, the photon energy spectrum is broadened
to a region with a width depending on $v_{s}$. This broadening effect
also manifests in our considered process: since the double-pole position
is determined by the real photon energy, a similar broadening of the
peak region is expected. 

\begin{figure}
\subfloat[$E_{s}=10\,\textrm{keV}$]{\includegraphics[width=0.33\textwidth]{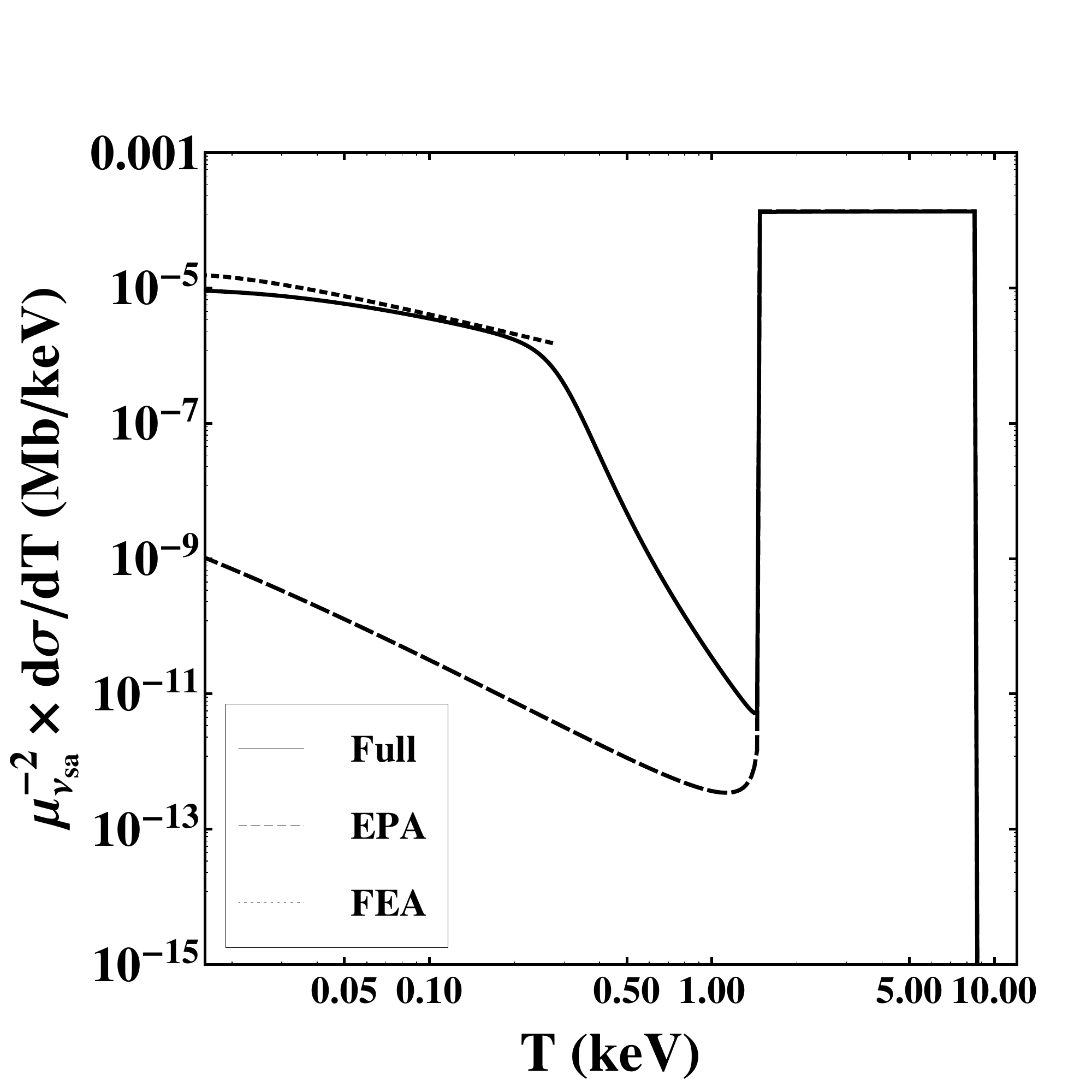}

\label{fig:nu_s-H_rel-1}}\subfloat[$E_{s}=100\,\textrm{keV}$]{\includegraphics[width=0.33\textwidth]{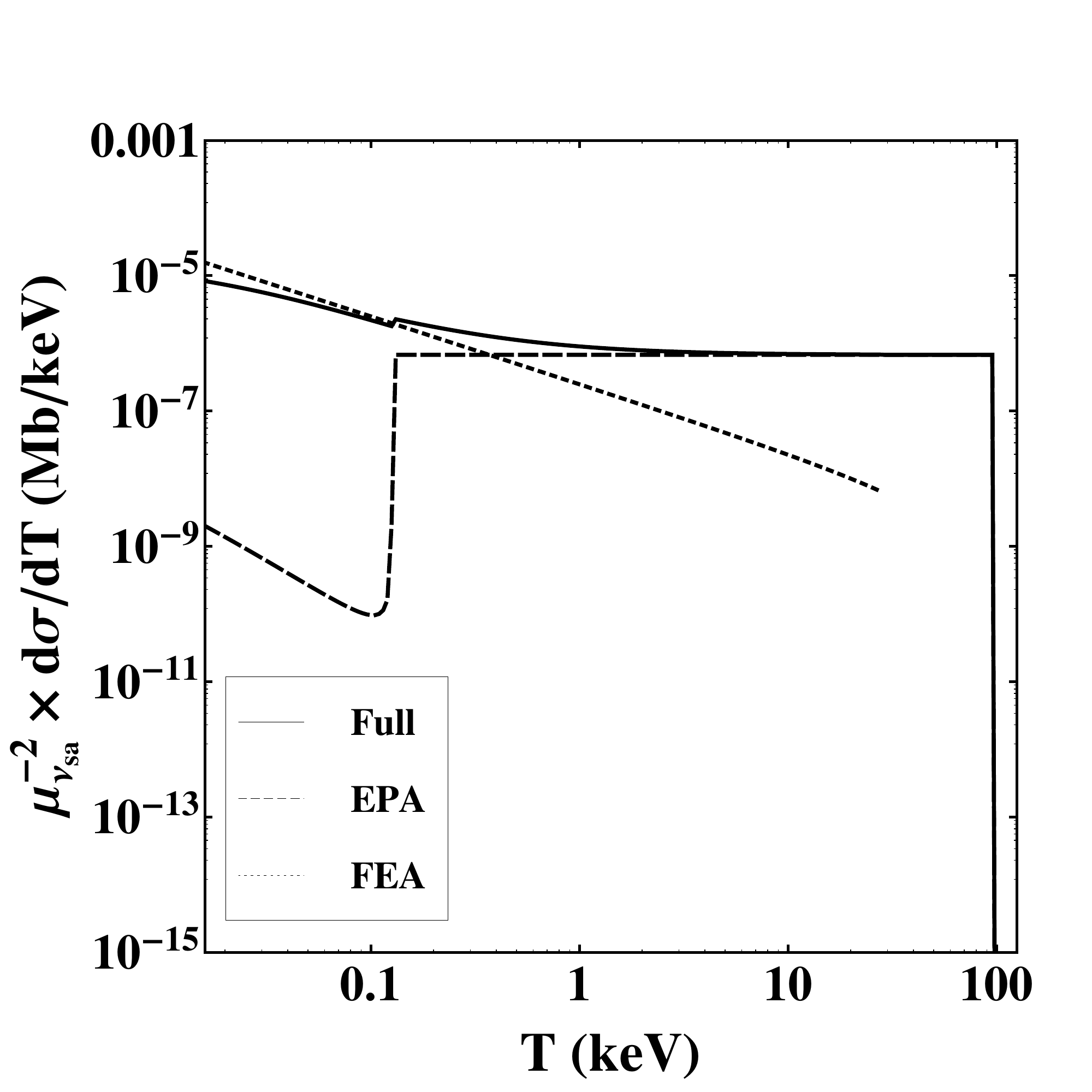}

\label{fig:nu_s-H_rel-2}}\subfloat[$E_{s}=1\,\textrm{MeV}$]{\includegraphics[width=0.33\textwidth]{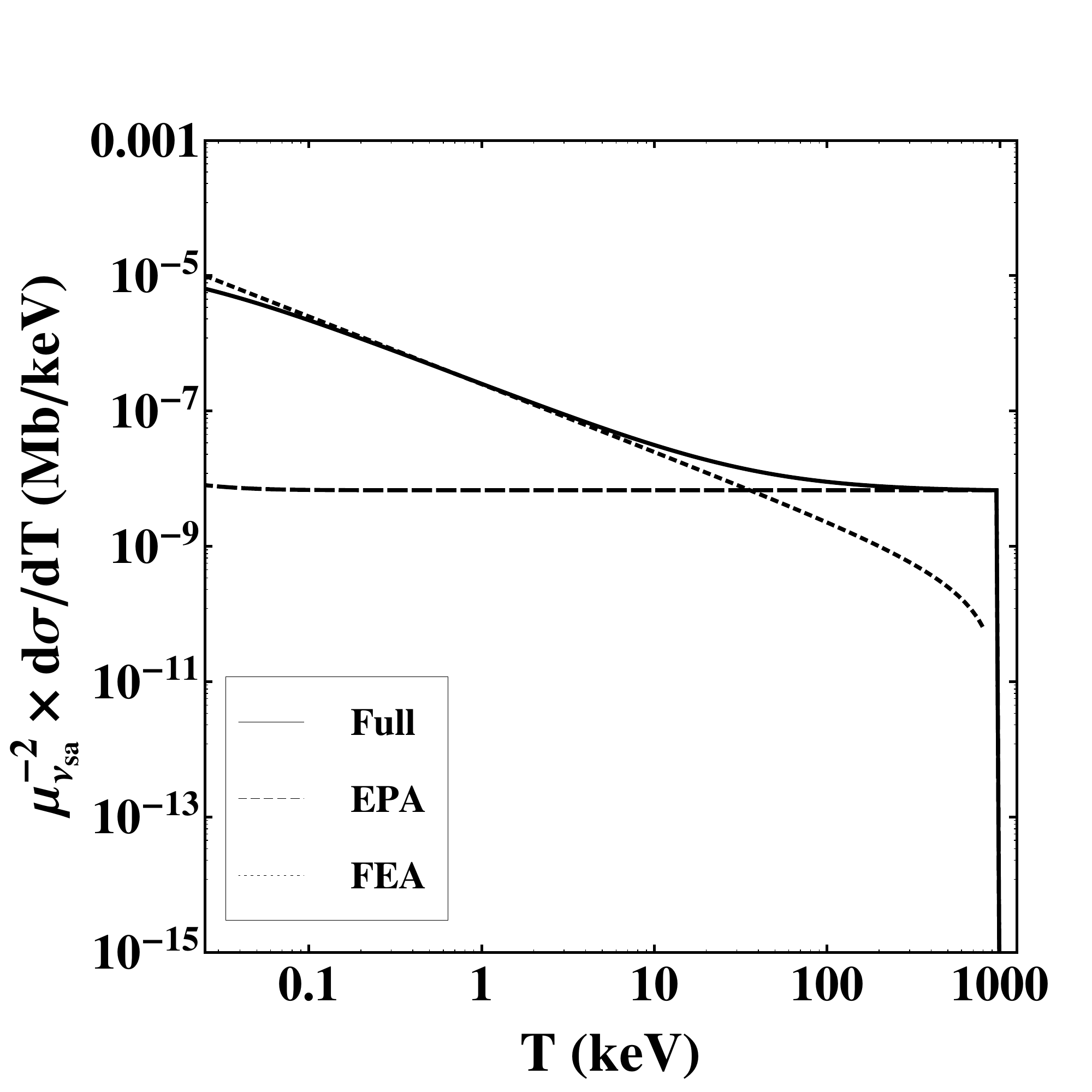}

\label{fig:nu_s-H_rel-3}}

\caption{The differential scattering cross sections of relativistic sterile
neutrinos ($m_{s}=7.1\,\textrm{keV}$) and hydrogen atoms through
the transition magnetic moment $\Ftwo$ with selected $E_{s}$. \label{fig:nu_s-H_rel}}
\end{figure}

The plots in Fig.~\ref{fig:nu_s-H_rel} show the results for relativistic
7.1-keV sterile neutrino with energy of (a) 10 keV, (b) 100 keV, and
(c) 1 MeV, respectively. Comparing these three cases, which differ
in relativistic degree, and the exact versus approximated results,
one can observe several important features:

(i) For $T$ above certain values till the end point, i.e., $E_{\nu}$,
the differential cross sections all behave like $T$-independent plateaux,
which can be well-described by the EPA. This plateau-like structure
is mainly due to the broadening of the double-pole peak and roughly
scales as $|\vec{k}_{s}|\sim E_{s}$, as explained in the previous
section. Also, because the EPA works well, the heights of these plateaux
and their scaling as $1/|\vec{k}_{s}|^{2}\sim1/E_{\nu}^{2}$ is anticipated
by Eq.~(\ref{eq:ds/dT_EPA_approx}). 

(ii) For smaller $T$, even when it still lies in the range where
a double pole is allowed kinematically, the EPA stops to be a good
approximation. This indicates that the longitudinal response starts
to contribute significantly. As can be seen from the figure, the exact
calculation overlaps less with the EPA plateau as the incident sterile
neutrino becomes more relativistic. 

(iii) In the medium to low $T$ region, on the contrary, the FEA becomes
a good approximation. In fact, with the incident sterile neutrinos
becoming more relativistic, it has a wider range of applicability.
For example, in the $E_{s}=1\,\textrm{MeV}$ case, the FEA works well
from near threshold all the way to $T\sim50\,\textrm{keV}$. Since
the mass of the sterile neutrino becomes negligible in the ultrarelativistic
limit, the differential cross section coverages to the one of neutrino
magnetic moment studies with active neutrinos. The latter case has
been extensively studied in hydrogen~\cite{Chen:2013iud} and complex
atoms such as germanium~\cite{Chen:2013lba}, all results show that
the FEA indeed is a good approximation for $T$ away from the threshold
and end point.

\section{Germanium case \label{sec:Germanium}}

Low threshold germanium detectors with sub-keV sensitivities have
played important roles in neutrino and dark matter experiments~\cite{Soma:2014zgm}.
In particular, they have been used to provide the stringent limits
on neutrino magnetic moments~\cite{Li:2002pn,Wong:2006nx,Beda:2012zz,Beda:2013mta}
and neutrino milli-charge~\cite{Chen:2014dsa}. The derivations formulated
in earlier sections are now extended to the germanium atom in this
section. 

The number density of germanium atoms in typical semiconductor detectors
is $n_{\textrm{Ge}}=4.42\times10^{22}/\textrm{cm}^{3}$, and this
gives the regulator $|\vec{q}|n_{\textrm{\textrm{H}}}\sigma_{\gamma}(\textrm{H},|\vec{q}|)\lesssim1200\,\textrm{eV}^{2}$
in the allowed range of $|\vec{q}|$ (the regulator decreases with
increasing $|\vec{q}|$). The atomic many-body physics is handled
by the multiconfiguration relativistic random phase approximation
(MCRRPA). The method has been benchmarked and applied to our previous
work on normal neutrino scattering through electroweak interactions.
Details can be found in Ref.~\cite{Chen:2014ypv}. 

\begin{figure}
\subfloat[$m_{s}=7.1\,\textrm{keV}$]{\includegraphics[width=0.45\textwidth]{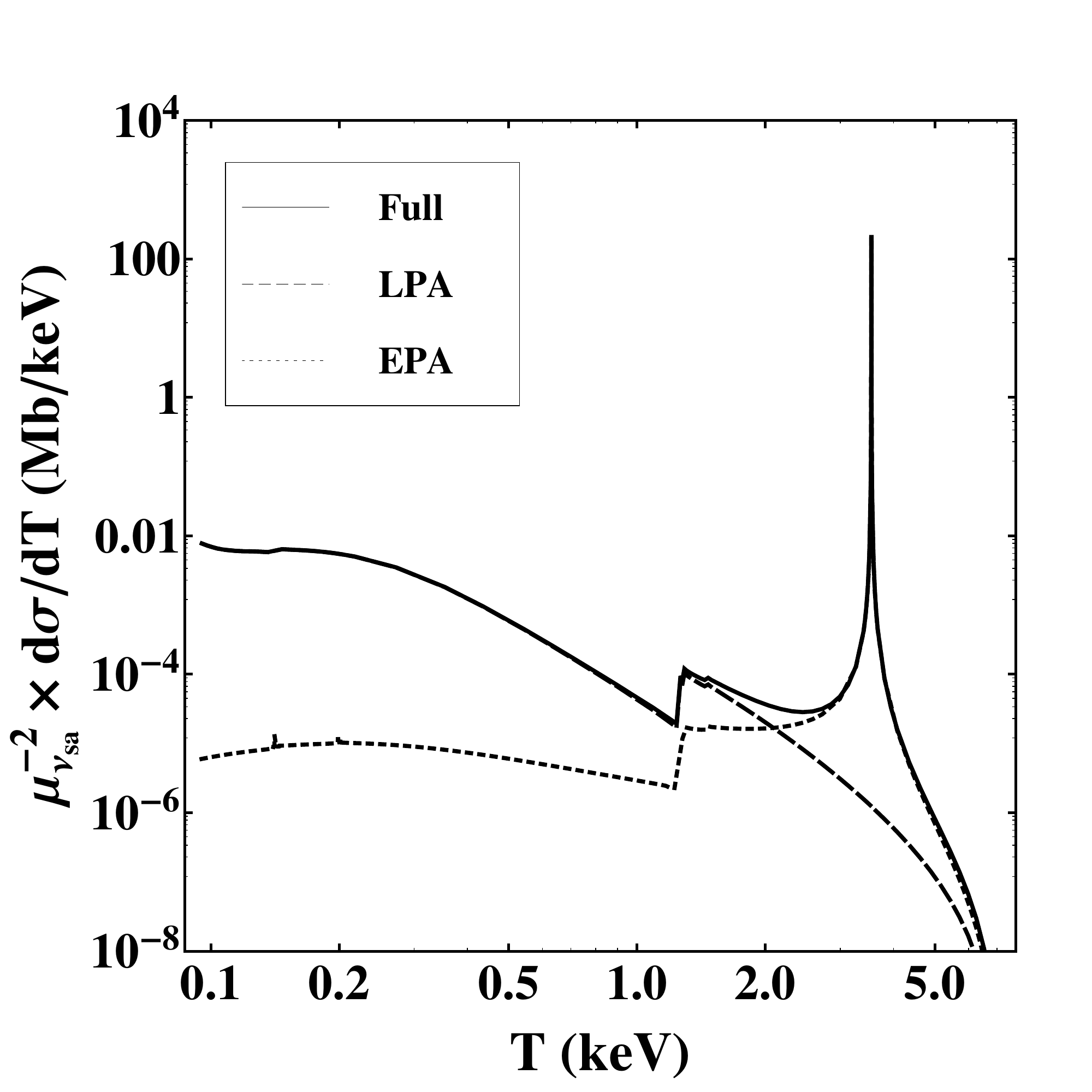}

\label{fig:nu_s-Ge_NR-1}}\subfloat[$m_{s}=20\,\textrm{keV}$]{\includegraphics[width=0.45\textwidth]{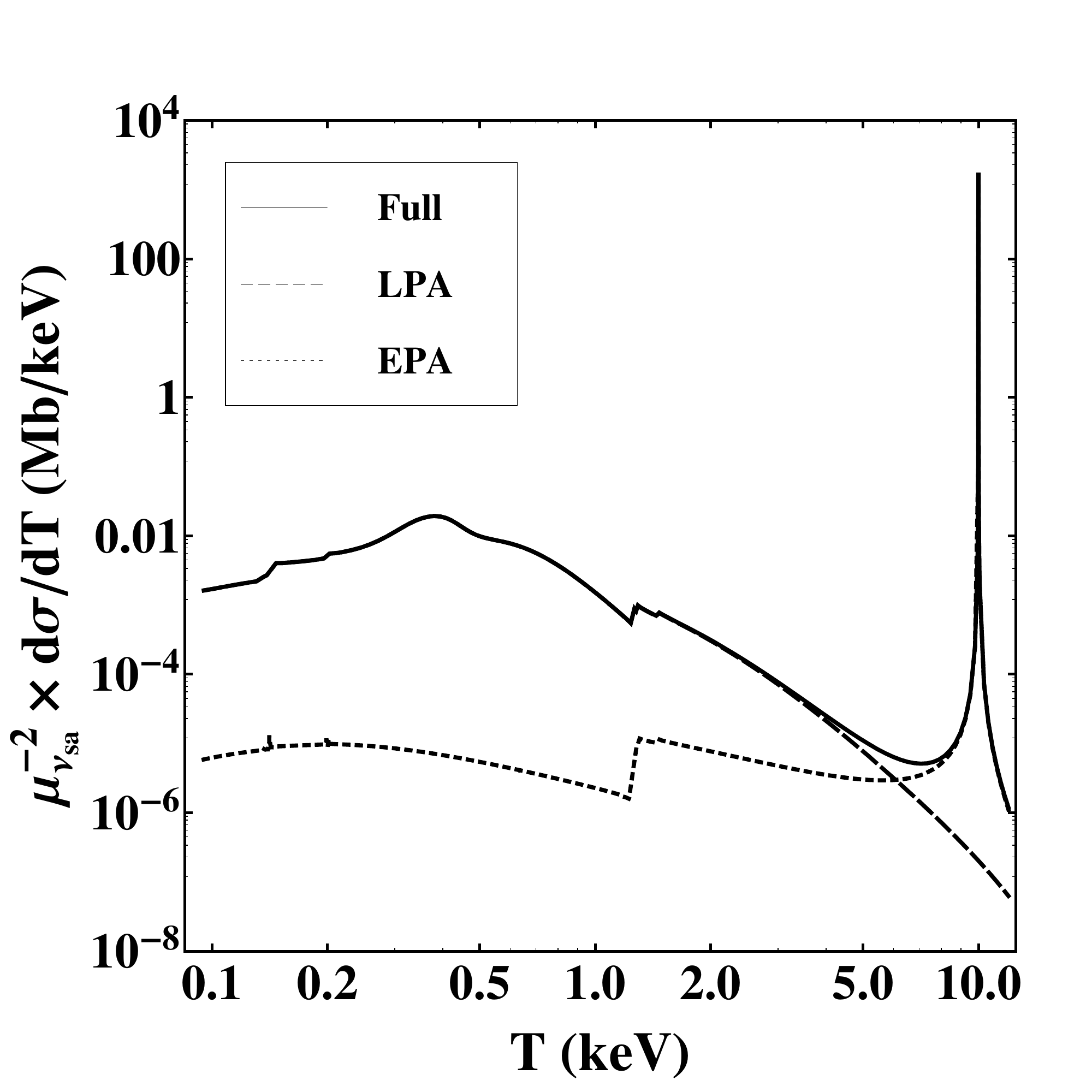}

\label{fig:nu_s-Ge_NR-2}}

\caption{The differential scattering cross section of nonrelativistic sterile
neutrinos ($v_{s}=10^{-3}$) and germanium atoms through the transition
magnetic moment $\Ftwo$ with selected $m_{s}$.\label{fig:nu_s-Ge_NR} }
\end{figure}

Fig.~\ref{fig:nu_s-Ge_NR-1} shows the results for the case: $m_{s}=7.1\,\textrm{keV}$
and $v_{s}=10^{-3}$. The peak region around $T=m_{s}/2=3.5\,\textrm{keV}$
to the end point is well approximated by the EPA, while at low recoil
energies, $T\lesssim1\,\textrm{keV}$, the LPA work better. In the
transition region between 1 keV and 3 keV, the transverse and longitudinal
responses contribute similarly in scale so neither approximations
are valid. The sharp edge observed at $T\sim1.3\,\textrm{keV}$ corresponds
to the opening of $n=2$ shells, which have ionization energies 1.26,
1.29, and 1.45 keV for $2p_{3/2}$, $2p_{1/2}$, and $2s_{1/2}$ orbitals,
respectively, as calculated by MCRRPA~\cite{Chen:2013lba,Chen:2014ypv}.
There are similar edges for higher orbitals (140, 145, and 202 eV
for $3p_{3/2}$, $3p_{1/2}$, and $3s_{1/2}$ orbitals, respectively),
however, not obvious on this log-log plot.

When $m_{s}$ is increased to 20 keV, with the same NR velocity, the
results are plotted in Fig.~\ref{fig:nu_s-Ge_NR-2}. The double-pole
peak is shifted to $T=m_{s}/2=10\,\textrm{keV}$ in the case, with
the peak value bigger than the $m_{s}=7.1\,\textrm{keV}$ case by
about one order of magnitude. This can be explained by the EPA formula,
Eq.~(\ref{eq:ds/dT_EPA_approx}), that $d\bar{\sigma}/dT\propto m_{s}^{2}$,
so $(20\,\textrm{keV}/7.1\,\textrm{keV})^{2}=10$. Various edges mentioned
previously are now resolved better in this plot. The peak at $T\sim370\,\textrm{eV}$
is the one corresponding to two-body $\nu_{s}$--$e$ scattering mentioned
previously, with the position predicted by Eq.~(\ref{eq:T_nu-e}).
{[}Note that the two-body peak for the $m_{s}=7.1\,\textrm{keV}$
case happens at $T\sim50\,\textrm{eV}$, which is outside the plot
range of Fig.~\ref{fig:nu_s-Ge_NR-1}.{]}

\begin{figure}
\includegraphics[width=0.6\textwidth]{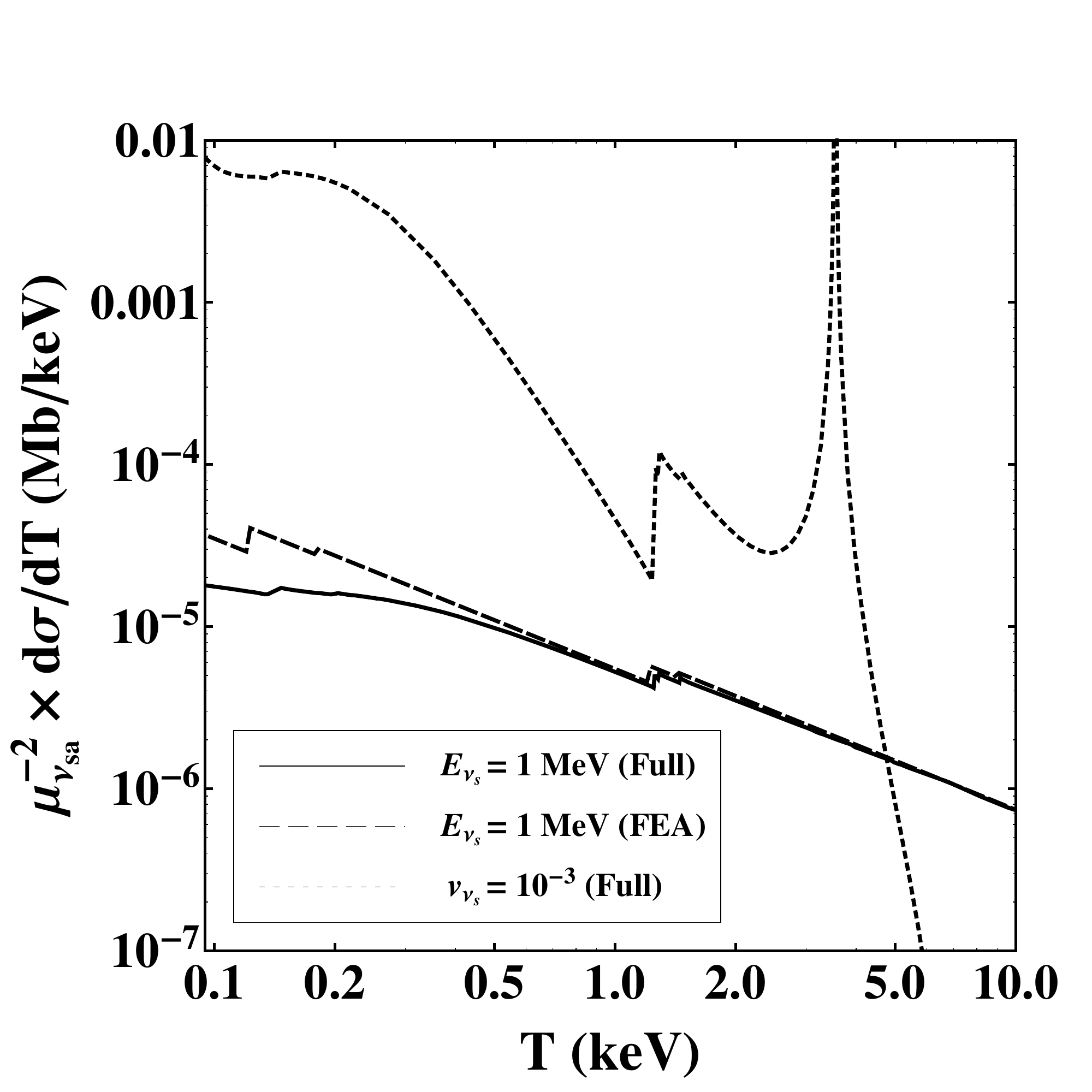}

\caption{The differential scattering cross section of relativistic sterile
neutrinos and germanium atoms through the transition magnetic moment
$\Ftwo$ with $m_{s}=7.1\,\textrm{keV}$ and $E_{\nu}=1\,\textrm{MeV}$
(solid line). For comparison, the results of free electron approximation
(dashed line) and the nonrelativistic case with $v_{s}=10^{-3}$ (dotted
line) are also shown. \label{fig:nu_s-Ge_rel} }
\end{figure}

Results for an ultrarelativistic 7.1-keV sterile neutrino of $E_{s}=1\,\textrm{MeV}$
are given in Fig.~\ref{fig:nu_s-Ge_rel}. For $T$ between 1 keV
to 10 keV, the FEA agrees with the MCRRPA result; for $T$ below 1
keV, the FEA slightly overshoots and differs from the MCRRPA result
by about a factor of 2 at $T=100\,\textrm{eV}$. Notice that these
two curves are almost identical to what have been shown in the upper
panel of Fig.~2 of Ref.~\cite{Chen:2013lba}, where neutrino magnetic
moments of active neutrinos were considered. In other words, the transition
magnetic moment arising from the sterile-to-active neutrino conversion
is indistinguishable from those from active neutrino mixings, as we
can take the zero mass limit for sterile and active neutrinos in a
relativistic process. In the same figure, we also compare the ultrarelativistic
and nonrelativistic 7.1-keV sterile neutrinos: In this low-recoil
regime being considered, $100\,\textrm{eV}\le T\le10\,\textrm{keV}$,
germanium detectors are more sensitive to the nonrelativistic sterile
neutrinos for they yield bigger differential cross sections in general
and exhibit rich, unique structure (can be resolved by detectors with
fine resolutions).

\section{Bounds on Sterile Neutrino Properties \label{sec:Bounds}}

\begin{figure}[!h]
\subfloat[\label{fig:bkgspectra-1}]{\includegraphics[width=0.45\textwidth]{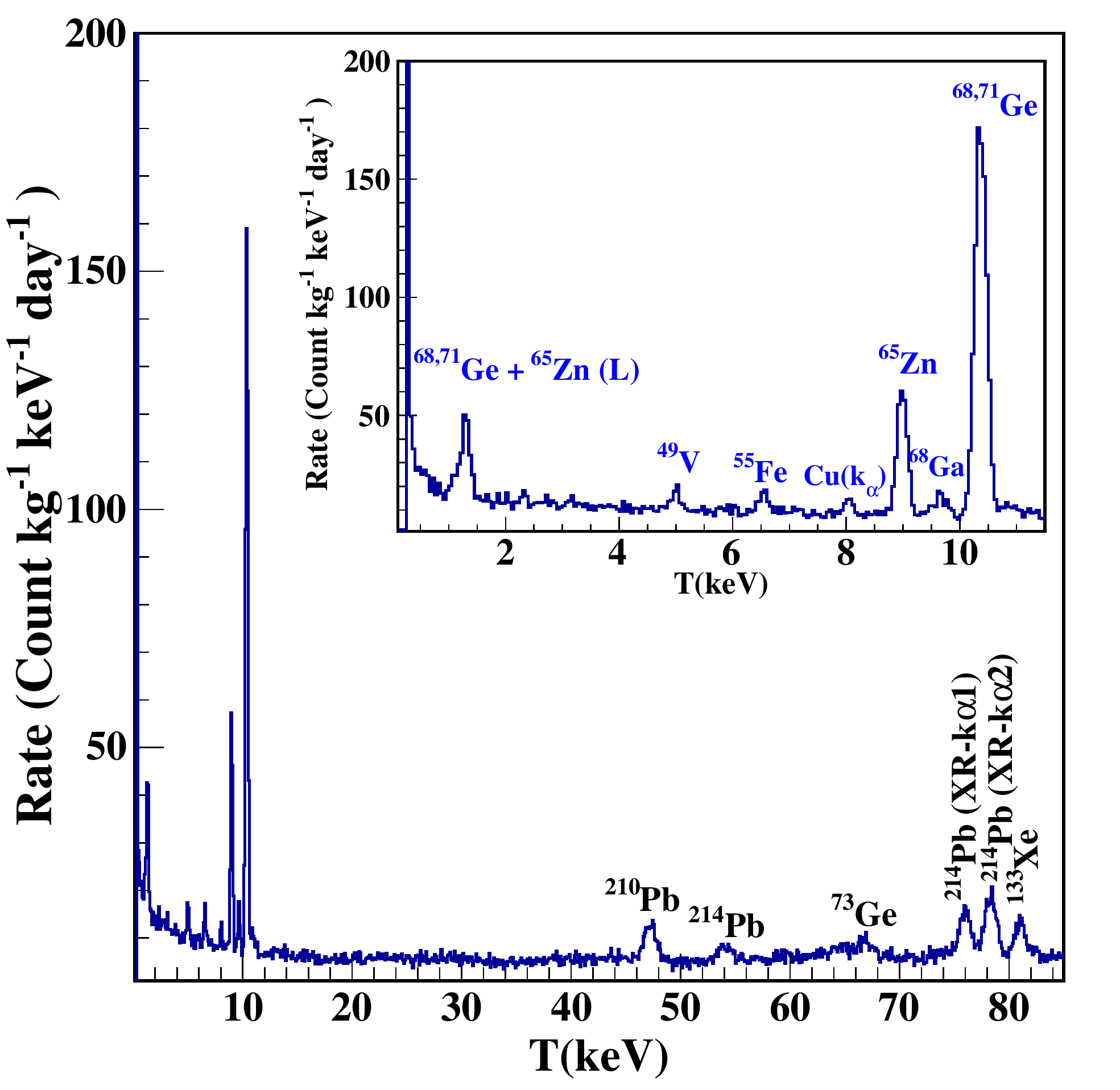}

} \subfloat[\label{fig:bkgspectra-2}]{\includegraphics[width=0.45\textwidth]{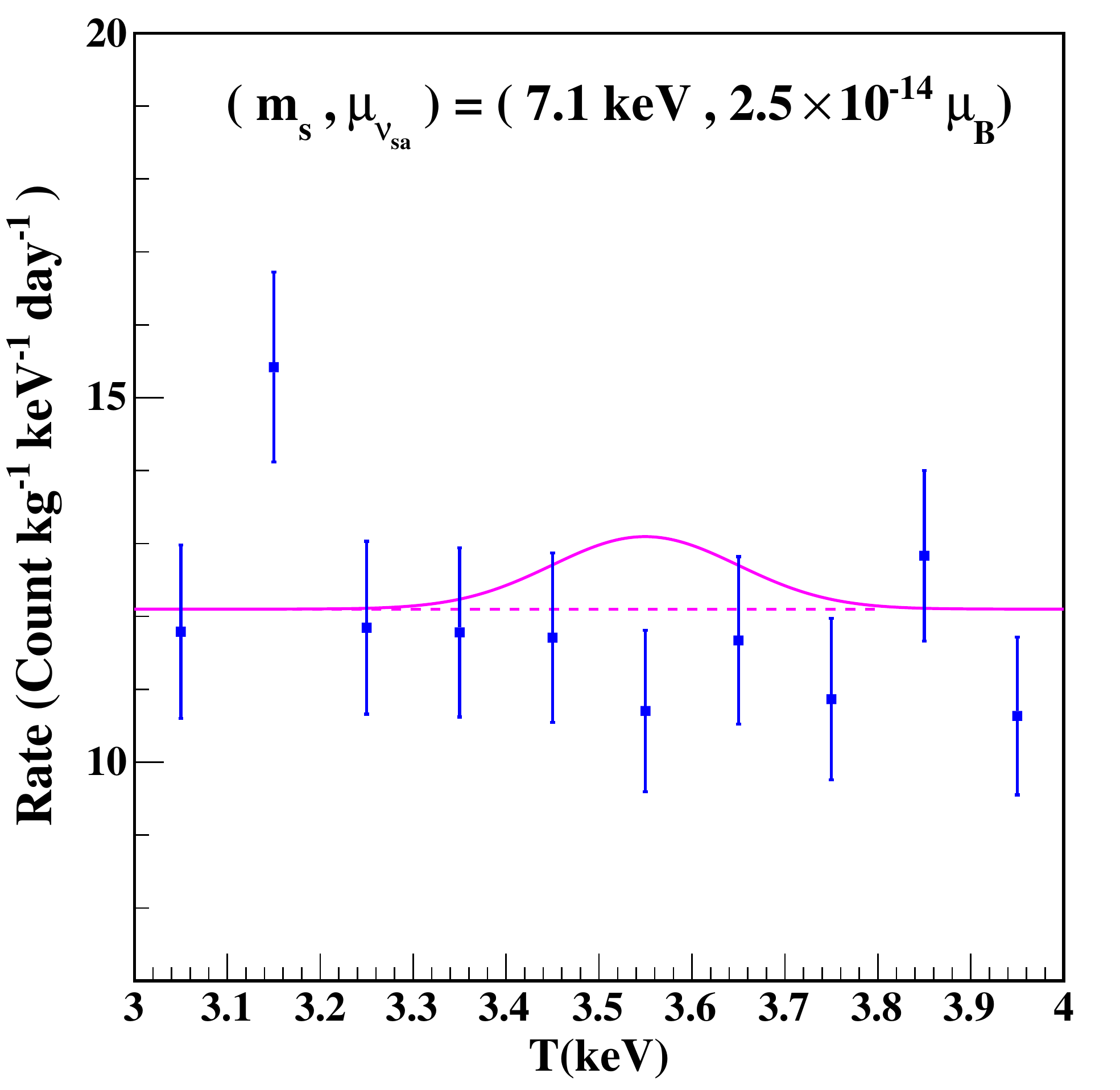}

} \caption{Experimental data and analysis: (a) Measured spectrum of germanium
detector at KSNL. All peaks can be accounted for by internal and ambient
radioactivity. (b) The zoomed energy range relevant to a $\nu_{s}$
with $m_{s}=7.1\,\textrm{keV}$. The spectrum due to $\mu_{\nu_{\textrm{sa}}}=2.5\times10^{-14}\,\mu_{\textrm{B}}$
excluded at 90\% CL is superimposed.}
\label{fig:bkgspectra} 
\end{figure}

A data sample of 139.3 kg-days with a 500 g n-type point contact germanium
detector taken at the Kuo-Sheng Reactor Neutrino Laboratory (KRNL)~\cite{Wong:2006nx,Deniz:2009mu}
were analyzed. The measured spectra after standard background suppression~\cite{Li:2002pn,Wong:2006nx,Deniz:2009mu}
is depicted in Fig.~\ref{fig:bkgspectra-1}. A dark matter analysis
searching for the atomic ionization interaction of Eq.~(\ref{eq::dmrate})
is applied to the data, using conventional astrophysical models on
the sterile neutrino as cold dark matter. 

The local dark matter density of $\rho$ = 0.4 GeVcm$^{-3}$ is adopted~\cite{Catena:2009mf}.
The event rate per unit mass on a target of germanium is given by
\begin{equation}
\left(\dfrac{dR}{dT}\right)=\dfrac{\rho_{s}}{m_{\textrm{A}}m_{s}}\int_{0}^{v_{\max}}\dfrac{d\sigma(m_{s},v)}{dT}vf(\vec{v})d^{3}v,\label{eq::dmrate}
\end{equation}
where $m_{\textrm{A}}$ is the mass of the germanium atom and $m_{s}$
denotes the mass of sterile neutrino. The normalized Maxwellian velocity
distribution
\begin{equation}
f(\vec{v})=N_{0}e^{(-\vec{v}^{2}/v_{0}^{2})}\Theta\left(v_{\textrm{esc}}-|\vec{v}|\right)\label{eq:maxvel}
\end{equation}
where $N_{0}$ is the normalization value and $\Theta$ denotes the
Heaviside step function. The dark matter particle has mean velocity
$v_{0}=220\,\textrm{km/s}$ and escape velocity $v_{\textrm{esc}}=533\,\textrm{km/s}$
in Earth's reference frame~\cite{Piffl:2013mla}.

\begin{figure}[!h]
\includegraphics[width=0.6\textwidth]{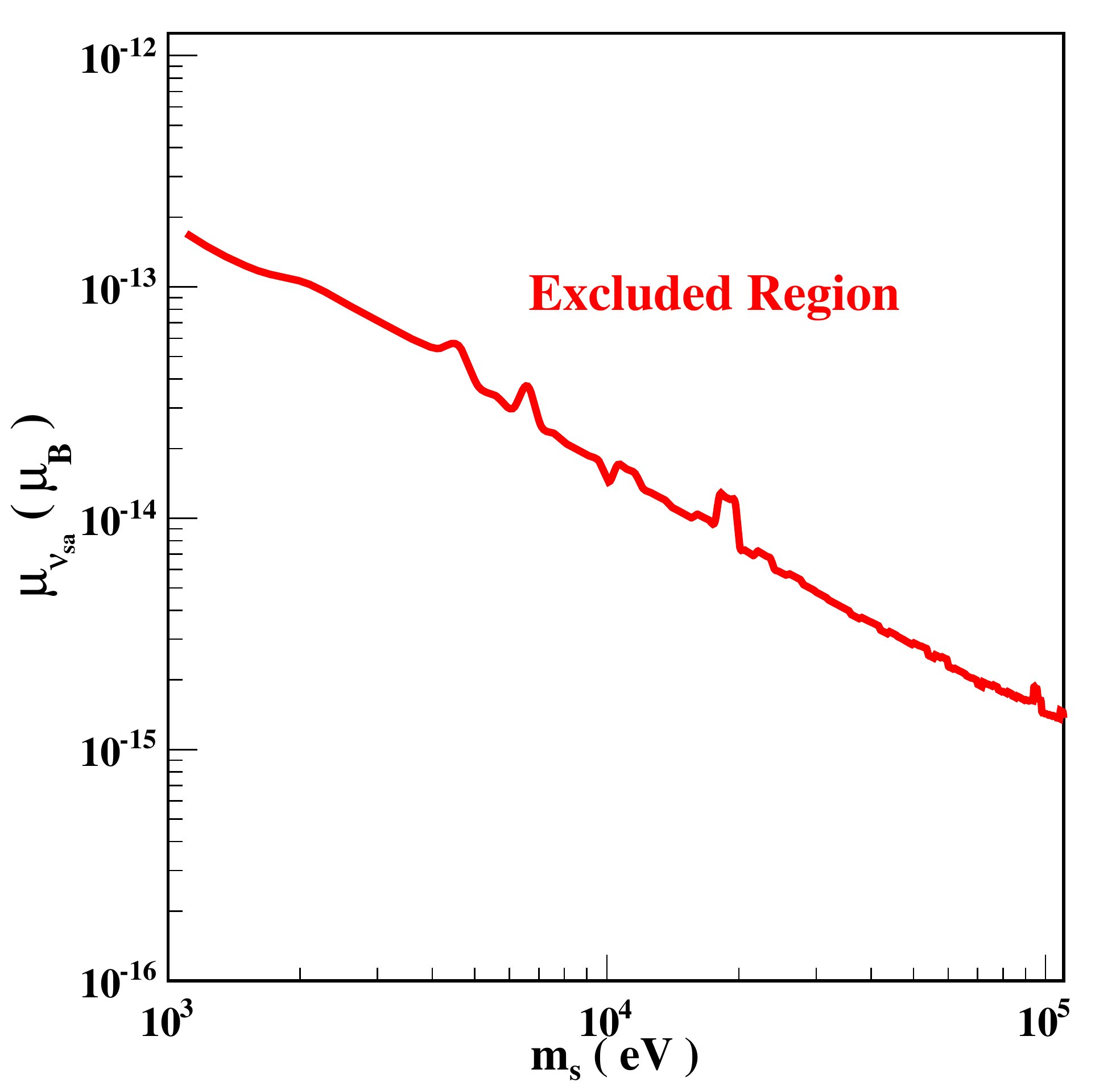} \caption{Exclusion curve at 90\% C.L. for the absolute value of the transition
magnetic moment of sterile neutrinos, based on reactor neutrino data
of Fig.~\ref{fig:bkgspectra}.}
\label{fig:exclplot} 
\end{figure}

Sterile neutrinos would manifest themselves experimentally as excess
of events over understood background with plateau-shaped spectrum
like that of Fig.~\ref{fig:bkgspectra-2}, due to convolution of
the theoretical signatures of Fig.~\ref{fig:nu_s-Ge_NR} and the
detector resolution. The data are analyzed by using the simplified
differential cross-section formula Eq.~(\ref{eq:ds/dT_EPA_approx}),
as the EPA provides a good approximation in this region. A minimal
$\chi^{2}$ analysis is applied with two free parameters describing
a locally smooth background and $\mu_{\nu_{\textrm{sa}}}$. As illustration,
the measurement in the vicinity $m_{s}=7.1\,\textrm{keV}$ is displayed
in Fig.~\ref{fig:bkgspectra-2} and the excluded spectrum at 90\%
CL is superimposed.

The exclusion plot of transition magnetic moment ($\mu_{\nu_{\textrm{sa}}}$)
versus mass ($m_{s}$) at 90\% C.L. is illustrated in Fig.~\ref{fig:exclplot}.
The bump structures in the exclusion correspond to the background
peaks from known radioactivity -- for instance, the drop in sensitivity
at $m_{s}\sim20\,\textrm{keV}$ is a consequence of increased background
due to the germanium X-ray peak at 10.37 keV ($K$-shell). At $m_{s}=7.1\,\textrm{keV}$,
the upper limit of $\mu_{\nu_{\textrm{sa}}}<2.5\times10^{-14}\,\mu_{\textrm{B}}$
at 90\% C.L. is derived.~\footnote{More precisely, it is the upper limit on the absolute value of $\mu_{\nu_{\textrm{sa}}}$,
as the experimental rate is proportional to $\mu_{\nu_{\textrm{sa}}}^{2}$.} As comparison, the laboratory upper limits of the magnetic moments
of $\nu_{e}$ and $\nu_{\bar{e}}$ are $1.3\times10^{-8}\,\mu_{\textrm{B}}$~\cite{Xin:2005ky}
and $2.9\times10^{-11}\,\mu_{\textrm{B}}$~\cite{Beda:2013mta},
respectively. The reason for a better sensitivity is our current case
is mainly due to the enhancement in the differential cross section.
We also note that the sterile neutrino DM flux on Earth at $m_{s}=7.1\,\textrm{keV}$
is of the same order of magnitude as the reactor electron anti-neutrino
flux at KSNL at a distance of 28 m from the reactor core. 

The radiative decay lifetime and transition magnetic moment of a sterile
neutrino can be related by Eq.~(\ref{eq:nu_s_decay}), so the recent
identification of a $7.1\,\textrm{keV}$ sterile neutrino with $\Gamma_{\nu_{s}\rightarrow\nu_{a}\gamma}=1.74\times10^{-28}\,\textrm{s}^{-1}$,
based on the astrophysical X-ray observations~\cite{Bulbul:2014sua,Boyarsky:2014jta},
can be converted to $\mu_{\nu_{\textrm{sa}}}=2.9\times10^{-21}\,\mu_{\textrm{B}}$.
This astrophysical determination exceeds our direct detection bounds
by several orders of magnitude, mainly because its much larger collecting
volume.

\section{Summary \label{sec:Summary}}

The transition magnetic moment of a sterile-to-active neutrino conversion
gives rise to not only radiative decay of a sterile neutrino, but
also its non-standard interaction (NSI) with matter. In this paper,
we consider the atomic ionization due to such a NSI, including hydrogen
and germanium. As the kinematics of this doubly inelastic scattering,
i.e., the projectile and target both change their internal states
(a massive to a massless neutrino for the former and a bound to a
free electron for the latter), can have a cross over between the space-like
and time-like regions in a certain range of energy transfer $T$,
the differential cross section is therefore enhanced whenever the
exchanged photon approaches the real limit. For a nonrelativistic
sterile neutrino with mass $m_{s}$ and velocity $v_{s}\ll1$, it
is found that the differential cross section exhibits a peak that
centers at $T\approx m_{s}/2$ with the width $\propto v_{s}$ and
maximum value $\propto m_{s}^{2}/v_{s}^{2}$. When the sterile neutrino
becomes more relativistic, the peak is smeared out due to the relativistic
broadening, so the transition magnetic moment of a sterile neutrino
is then indistinguishable from the ones of active neutrinos.

Using the data taken by the TEXONO germanium detectors, which have
fine energy resolution in keV and sub-keV regimes, we derive constraints
on the mass and transition magnetic moment $\mu_{\nu_{\textrm{sa}}}$
of a sterile neutrino as the dark matter particle. For $m_{s}$ in
the range of a few keV to 100 keV, the upper limit on $\mu_{\nu_{\textrm{sa}}}$
drops from $\sim10^{-13}\,\mu_{\textrm{B}}$ to $\sim10^{-15}\mu_{\textrm{B}}$
with increasing $m_{s}$. These constraints are better than the current
direct limits on the magnetic moments of active neutrinos, $\sim10^{-11}\,\mu_{\textrm{B}}$,
mainly because of the much enhanced scattering cross sections at $T\approx m_{s}/2$.
On the other hand, the astrophysical hints of a 7.1-keV sterile neutrino
with radiative decay rate $\Gamma_{\nu_{s}\rightarrow\nu_{a}\gamma}=1.74\times10^{-28}\,\textrm{s}^{-1}$
would imply a more sensitive determination of $\mu_{\nu_{\textrm{sa}}}=2.9\times10^{-21}\,\mu_{\textrm{B}}$,
due to the huge collecting volume of galactic scales.
\begin{acknowledgments}
We acknowledge the support from the Ministry of Science and Technology,
Taiwan under Grants Nos. 102-2112-M-002-013-MY3 (J.-W. C., C.-L. W.,
and C.-P. W.), 103-2112-M-259-003 and 104-2112-M-259-004-MY3 (C.-P.
L.), 104-2112-M-001-038-MY3 (H. T. W. and L. S.); the Center for Theoretical
Sciences and Center of Advanced Study in Theoretical Sciences of National
Taiwan University (J.-W. C., C.-L. W., and C.-P. W.); and the National
Center for Theoretical Sciences. 

J.-W. C would like to thank the hospitality of the Rudolph Peierls
Centre for Theoretical Physics of the University of Oxford and Oxford
Holography group, DAMTP of University of Cambridge, and Helmholtz-Institut
für Strahlen-und Kernphysik and Bethe Center for Theoretical Physics,
Universität Bonn.
\end{acknowledgments}

\bibliographystyle{apsrev4-1}
\bibliography{draft}

\end{document}